\documentclass[journal]{IEEEtran}
\usepackage{graphicx}
\usepackage{subfigure}
\usepackage{amsmath}
\usepackage{multirow}

\begin{document}

\title{Analysis and Modeling of Network Connectivity in Routing Protocols for MANETs and VANETs }

\author{S. N. Mohammad$^{1}$, M. J. Ashraf$^{2}$, S. Wasiq$^{1}$, S. Iqbal$^{3}$, N. Javaid$^{1}$\\\vspace{0.4cm}
        $^{1}$COMSATS Institute of Information Technology, Islamabad, Pakistan.\\
        $^{2}$Islamabad Electric Supply Corporation, Islamabad, Pakistan.\\
        $^{3}$Abasyn University, Peshawar, Pakistan.\\

     }

\maketitle

\begin{abstract}
In this paper, a framework is presented for node distribution with respect to density, network connectivity and communication time. According to modeled framework we evaluate and compare the performance of three routing protocols; Ad-hoc On-demand Distance Vector (AODV), Dynamic Source Routing (DSR) and Fisheye State Routing (FSR) in MANETs and VANETs using two Mac-layer protocols; 802.11 and 802.11p. We have further modified these protocols by changing their routing information exchange intervals; MOD AODV, MOD DSR and MOD FSR. A comprehensive simulation work is performed in NS-2 for the comparison of these routing protocols for varying mobilities and scalabilities of nodes. To evaluate their efficiency; throughput, End-to-End Delay (E2ED) and Normalized Routing Load (NRL) of these protocols are taken into account as performance parameters. After extensive simulations, we observe that AODV outperforms both with MANETs and VANETs.
\end{abstract}


\IEEEpeerreviewmaketitle

\section{Background and Motivation for the Work}
Routing protocols are designed to calculate paths for communication networks. The routing protocols are subdivided into table driven, on-demand and hybrid based on the topology of the network. In table driven, proactive protocols are based upon periodic exchange of control messages and maintains routing tables. Each node maintains complete information about the network topology locally. However, the reactive protocol tries to discover a route only when demand arrives. It usually takes more time to find a route as compare to a proactive protocol.

Our simulation work based upon three protocols comparison in MANETs and in VANETs named as AODV [1], DSR [2] and FSR [3]. The table. 1 demonstrates the comparison of these three routing protocols with respect to their routing strategies in detail. Moreover, we introduce some modifications in their routing exchange intervals; 1) in MOD AODV, augments AODV's Expanding Ring Search algorithm (ERS) limits, 2) in MOD DSR, time associated with storage of routes in Route Cache is modified and 3) Scope intervals in FSR are adjusted in MOD FSR.

In [4] [5], communication time between nodes is found when the nodes are moving in same and opposite direction with same or different speeds. In this study, we improve the work of [6-10] and calculate the probability of link establishment between nodes when they are moving in same and opposite direction with same or different speeds. The study of DYMO is done in comparison with other routing protocols [6]. The performance metrics such as jitter, throughput and delay is taken under observation in their work.

The study [6] involved the consistently varying network topology and comparison of DSR with AODV in MANETs for different scenarios and performance metrics to propose the best scenario for each routing protocol to maximize its efficiency. In [7] the authors modified the OLSR protocols in their paper. We also do some modifications and evaluate AODV, DSR and FSR for both MANETs and VANETs.

The comparison for AODV and DSR in VANETs is carried out for realistic urban scenario with variable node mobility and vehicle density to observe the behavior of both protocols [5].




\section{Modeled Mathematical Framework}
\subsection{Distribution of Node Population Size}

Each stream of nodes may be modeled as M/E/$\infty$ queue with the service time of a node being the amount of time that it spends in the service strip. Thus the service time of a node terminates with its departure from the service strip.
This work determine the steady-state distribution of the number of nodes within each segment. Let $N_{kj}$ is Poisson distribution with the parameter $\tilde{\lambda}_{kj}\int_{0}^{t}B_{x(\tau)}(R_{kj})d\tau$. The corresponding steady state distribution can be found by determining the limit of the Poisson parameter as the time approaches to infinity.
Let ${N_{i}}$ denote the node population within segment $i$. Defining $k$ as the sub-strip that segment $i$ is located, then $k={max(1,....j...k)}{\forall}{i_{j}<i}$.
$N_{i}$ has also Poisson distribution with parameter $\phi_{i}$.

\small
\begin{eqnarray}
\phi_{kj}=\lim_{t\rightarrow\infty}\tilde{\lambda}_{kj}\int_{0}^{t}B_{x(\tau)}(R_{kj})d\tau
\\\phi_{kj}=\lim_{t\rightarrow\infty}\tilde{\lambda}_{kj}\int_{0}^{t}\int_{0}^{R_{kj}}\frac{1}{\sqrt{2\pi\theta_{x(\tau)}}}e^{-\frac{(y-\mu\tau)^2}{2\theta_{x(\tau)}}}dyd\tau
\end{eqnarray}
\normalsize

This integral may be evaluated numerically for large values of $t$.

\begin{eqnarray}
\phi_{i}=\sum_{\ell=ln=k+1}^{k}\sum_{\ell=ln=k+1}^{K}\tilde{\phi}_{\ell{n}}(i)
\end{eqnarray}

where, $ \tilde{\phi}_{\ell{n}}(i)=\phi_{\ell{n}}P_{\ell{n}}(i)$ and;

$P_{\ell}$(i)=$Prob(a node from S_{ln} stream is located in the segment i)$

\tiny
\begin{eqnarray}
P_{\ell{n}}(i)=\int_{(i-1)d-R_{1i}}^{id-R_{1i}}\tilde{b}_{x_{\ell{n}}}(r)dr=\frac{e^{mid}-e^{m(i-1)d}}{e^{mR_{1i}}(e^{mR_{\ell{n}}}-1)}
\end{eqnarray}
\normalsize
The probability distribution of the number of the nodes within segment $i$ and its probability generating function (PGF) at the steady state is given by:
\small
\begin{eqnarray}
Pr(\tilde{N}_{i}=n)=e^{-{\phi}_{i}}\frac{\phi_{i}^{n}}{n!}\,\,\,\,\,\ and \,\,\,\,\,P(z)=E[z^{N_{i}}]=e^{-\phi_{i}(1-z)}
\end{eqnarray}
\normalsize

Fig.1 shows detail scenario of communication time between node with probability of link establishment. It also shows probability of node communication.


The above mention work can be extended and improvement can be made in the distribution of node population size. Probability of node population was carried out within a segment $N_{i}$. The improvement has been made to find the node population size within the whole topology containing $n$ segments. Let $N_{n_{total}}$ denote the population within $n$ segments. Defining $s$ as the strip containing $n$ segments, then the Poisson distribution of $N_{n_{total}}$ with parameter $\phi_{n_{total}}$ is given by:
\small
\begin{eqnarray}
\phi_{n_{total}}=\sum_{\ell=1}^{s}\sum_{\ell=s+1}^{S}\tilde{\phi}_{\ell{n}}(n_{total})
\end{eqnarray}
\normalsize
where, $ \tilde{\phi}_{\ell{n}}(n_{total})=\phi_{\ell{n}}P_{\ell{n}}(n_{total})$ and;

\tiny
\begin{eqnarray}
\begin{split}
P_{\ell{n}}(n_{total})=\int_{(n_{total}-1)d-R_{1n_{total}}}^{n_{total}d-R_{1n_{total}}}\tilde{b}_{x_{\ell{n}}}(r)dr \\=\frac{e^{mn_{total}d} - e^{m(n_{total}-1)d}}{e^{mR_{1n_{total}}}(e^{mR_{\ell{n}}}-1)}
\end{split}
\end{eqnarray}
\normalsize
Where $d$ is node's transmission range, $R$ is total length of the strip containing $n$ segments.
Also;\\
$m=\frac{\mu\beta}{2\sigma^{2}+\mu^2}$
The probability distribution of the number of the nodes within $n$ segments and its PGF at the steady state are given by:
\tiny
\begin{eqnarray}
\begin{split}
Pr(\tilde{N}_{n_{total}}=n)=\sum_{T=1}^{n_{total}}e^{-{\phi}_{n_{total}}}\frac{\phi_{n_{total}}^{n}}{n!} \\ P(z)=E[z^{N_{n_{total}}}]=\sum_{T=1}^{n_{total}}e^{-\phi_{n_{total}}(1-z)}
\end{split}
\end{eqnarray}
\normalsize

\subsection{Analysis of Network Connectivity}
In this section, we will determine the network connectivity of a new arriving node at the beginning of the service strip at the steady state.
The new arrival node will be designated as the cluster head and all nodes within the cluster will have direct or indirect communications with the new arrived node. It is assumed that two nodes will be able to communicate directly if $L<d$ where $L$ is the distance between the nodes and $d$ is the constant transmission range of a node. Clearly, all the nodes within a given segment are able to communicate directly as they are within each other's transmission range. The new arriving node will see the equilibrium distribution of population size and it will also be able to communicate directly with all the nodes within the first segment.

Next, let us define the direct communication probability of two nodes at consecutive segments $i$ and $i+1$ as;

\small
$P_{i}$=$Pr(L<d)$ two nodes are located at consecutive segments $i \hspace{0.1cm}and\hspace{0.1cm} $i+1)
\normalsize

Moreover this probability of two nodes located within two consecutive segments has a constant value of $ 1/2 $.
Let random variable  $ x_{i}  $ denote the distance of a node located at the segment $i$ from the beginning of the service strip and $ b_{x_{i}(r)}$  the corresponding pdf of this variable. A node in this segment may belong to any of the first $k$ streams. Thus $ b_{x_{i}(r)} $ will be determined by weighted average of the pdfs of distances of nodes from different streams given by $ \tilde{b}_{x_{kj}(r)}$. We also have to take into consideration that pdf of the distance for each stream is relative to its service point. Thus;

\hspace{-0.5cm}
\tiny
\begin{align}
b_{x_{i}(r)}&=\frac{1}{\phi _i}\sum_{\ell =ln=k+1}^{k}\sum_{\ell  \nonumber =ln=k+1}^{K}\tilde{\phi}_{\ell{n}}(i)\frac{\tilde{b}_{x_{\ell{n}}}(r-R_{1{\ell}})}{\tilde{B}_{x_{\ell{n}}}(id-R1{\ell}) -\tilde{B}_{x_{\ell{n}}}((i-1)d-R{1{\ell}})}\\
&=\frac{1}{\phi_{i}}\sum_{\ell  \nonumber =ln=k+1}^{k}\sum_{\ell=ln=k+1}^{K}\tilde{\phi}_{\ell{n}}(i)\frac{me^{m(r-R1{\ell})}}{e^m(id-R1{\ell})-e^m((i-1)d-R1{\ell})}\\ \nonumber
&=\frac{1}{\phi_{i}}\sum_{\ell =ln=k+1}^{k}\sum_{\ell =ln=k+1}^{K}\tilde{\phi}_{\ell{n}}(i)\\\nonumber
&=\frac{me^{mr}}{e^{imd}-e^{(i-1)md}}\frac{1}{\phi_{i}}\sum_{\ell =ln=k+1}^{k}\sum_{\ell =ln=k+1}^{K}\tilde{\phi}_{\ell{n}}(i)\\
&=\frac{me^{mr}}{e^{imd}-e^{(i-1)md}}
\end{align}
\normalsize
where, $ (i-1)d\le{x_{i}}\le{id}, 1\le{i}\le{I-1}$ and $m=\frac{\mu\beta}{2\sigma^2+\mu^2}$

The direct communication probability of two nodes located at consecutive segments $i$ and $i+1$ respectively is given by,
\tiny
\begin{eqnarray}
P_{i}=Pr(x_{i+1}-x_{i}<d)=\int_{o}^{r_{i}}\int_{r_{i}}^{r_{i+1}}b_{x_{i}}(r_{i})b_{x_{i+1}}
(r_{i+1})dr_{i}dr_{i+1}
\end{eqnarray}
\normalsize


\begin{figure*}[t]
\centering
  {\includegraphics[height=9 cm,width=12 cm]{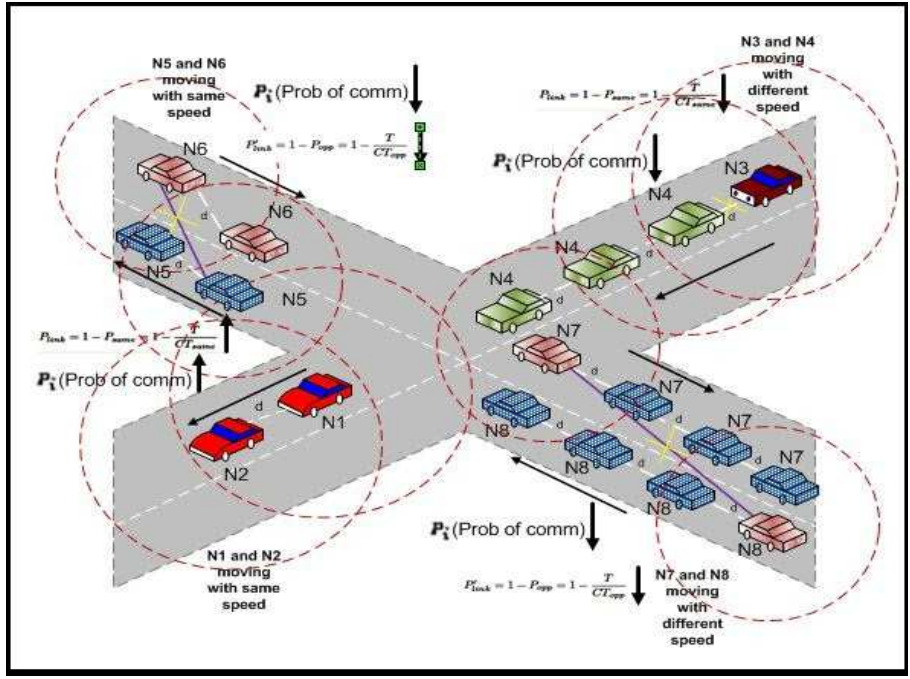}}
  \caption{System Model}
\end{figure*}

\subsection{Indirect Communication}
The probability of direct communication is given in eq.(5). For three segments, the probability of nodes having indirect communication between them is given by:
\tiny
\begin{eqnarray}
\begin{split}
P_{i+1}=Pr(x_{i+2}-x_{i+1}<d)
\\=\int_{r_{i}}^{r_{i+1}}\int_{r_{i+1}}^{r_{i+2}}b_{x_{i+1}}(r_{i+1})b_{x_{i+2 }}(r_{i+2})dr_{i+1}dr_{i+2}
\end{split}
\end{eqnarray}
\normalsize
Now, by combining eq.(5) and eq.(6), we can find the total probability of indirect communication between three segments as:
\begin{eqnarray}
P_{i}''=P_{i}\times P_{i+1}
\end{eqnarray}

For finding the efficiency of eq.(7);

\begin{eqnarray}
E_{i}=P_{i}'' \times 100\%
\end{eqnarray}

Similarly, for indirect communication of nodes in $n$ segments we write:
\begin{eqnarray}
P_{n_{total}}= \prod_{i=1}^{n_{total}-1}P_{i} \times P_{i+1}
\end{eqnarray}



As the number of nodes increases in each segment the probability of them having direct communication decreases. The decrease in probability is eminent from the fact that as the node density increases, the interference between nodes increases and each node cannot directly communicate with one another.
The probability of communication increases only when the segment is not congested with large number of nodes and each node falls in the transmission range of another node.
Efficiency of nodes having indirect communication in $n$ segments can be found as: $E_{n_{total}}=P_{n_{total}} \times 100\%$. As we mentioned earlier that probability of node communication can be seen from Fig. 2 as well. The Probability of direct communication of two segments given in [8] is 1/2, we can generalized it as: $p_{n_{total}}=\frac{1}{n}$ where $n$ is the node density and $ P(n_{total}) $ is the corresponding probability.
The effects can be seen in Fig. 2.


\begin{figure}[h]
\begin{center}
\includegraphics[height=6 cm,width=8 cm]{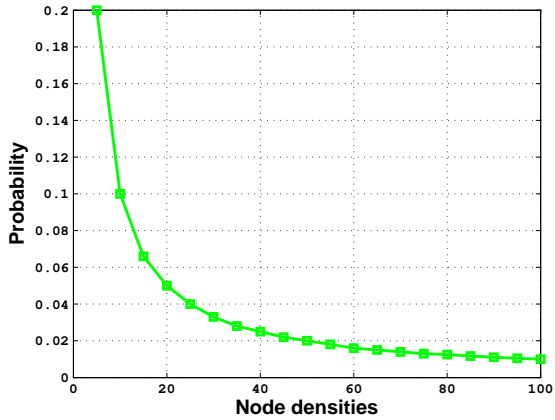}
\caption{Probability of Communication}
\end{center}
\end{figure}
\vspace{-0.5cm}

\subsection{Modeling Of Communication Time}
In the previous sections we discussed about the probability of node population within a given segment and direct as well as indirect communication of nodes.
We can further extend our work on the communication of nodes by calculating the communication time for which the given set of nodes will have a direct communication within respective segments. The route breakage probability of nodes within the segments is modeled in [4], as the probability is independent of time.

The average link breakage probability between nodes moving in the same direction $ P_{same}$ can be found as:

\begin{eqnarray}
P_{same} = \frac{T}{CT_{same}}
\end{eqnarray}

Where $T$ is the simulation time and $ CT_{same} $ is the average total time of node communication.

Let $ d=R/n $ be the separation distance between the nodes, $R$ is the total length of the strip and $n$ is the total number of nodes on the strip [5]. The total number of nodes in a segment $i$ can be given as $ N_{i} = \frac{T_{r}}{d} $, where $ T_{r} $ is the nodes transmission range. The average lifetime of the links among the nodes that form route is given by:

\begin{eqnarray}
CT_{same} = \frac{T_{same_{s}} + T_{diff_{s}}}{(V_{c}+1)^2}
\end{eqnarray}

Where, $ T_{same} $ is the communication time of the nodes moving in the same direction at the same speed and $ T_{diff}$ is the communication time of the nodes moving in the same direction with different speed: Also;
\begin{eqnarray}
V_{c}=\frac{V_{max}-V_{min}}{\Delta{V}}
\end{eqnarray}

Where, $V_{c}$ is the speed of nodes different speed and $\Delta{V}$ is the change in speed of nodes which is taken into calculation. $V_{max}$ and $V_{min}$ are the maximum and minimum speed at which nodes are moving respectively. Now we can find the average communication time $ T_{comm_{same}}$ between nodes moving in the same direction at same speed as:

\begin{eqnarray}
T_{comm_{same}}=\frac{\sum_{i=j=0}^{V_{c}}\sum_{m=1}^{N_{i}}T}{N_{i}}
\end{eqnarray}

$ T_{comm_{diff_{s}}}$ is the average communication time between the nodes moving in the same direction but with different speeds, which can be calculated as:

\tiny
\begin{eqnarray}
T_{comm_{diff}}=\sum_{i=0}^{V_{c}}\sum_{j=0}^{V_{c}}\frac{\sum_{m=1}^{N_{i}}\frac{T_{r}-md}{|{(V_{min}+i\Delta{V})-(V_{min}+j\Delta{V})}|}}{N_{i}}
\end{eqnarray}

\normalsize
The total probability of communication between nodes moving in the same direction, when link is established is denoted as $P_{link}$ and is given as:
\begin{eqnarray}
P_{link}=1-P_{same}=1-\frac{T}{CT_{same}}
\end{eqnarray}

The probability of indirect communication for three segments is shown in eq.(7), we relate it with the probability of route establishment between the nodes given in eq.(15) as:

\tiny
\begin{eqnarray}
\begin{split}
P_{i}'=Pr(x_{i+2}-x_{i+1}<d)\\
=\int_{r_{i}}^{r_{i+1}}\int_{r_{i+1}}^{r_{i+2}}b_{x_{i+1}}(r_{i+1})b_{x_{i+2 }}(r_{i+2})dr_{i+1}dr_{i+2} \times P_{link}
\end{split}
\end{eqnarray}
\normalsize
or
\tiny
\begin{eqnarray}
\begin{split}
P_{i}'=Pr(x_{i+2}-x_{i+1}<d)
\\=\int_{r_{i}}^{r_{i+1}}\int_{r_{i+1}}^{r_{i+2}}b_{x_{i+1}}(r_{i+1})b_{x_{i+2 }}(r_{i+2})dr_{i+1}dr_{i+2} \times 
\end{split}
\end{eqnarray}
\normalsize
Similarly,

\begin{eqnarray}
P_{i}''=P_{i}\times P_{i}' \times P_{link}=P_{i}\times P_{i}' \times (1-\frac{T}{CT_{same}})
\end{eqnarray}

Now the probability of indirect communication of $n$ shown  in eq.(9) can be transformed as:
\begin{eqnarray}
P_{n_{total}}= \prod_{i=1}^{n_{total}-1}P_{i}\times P_{i+1}\times P_{link}
\end{eqnarray}
or
\begin{eqnarray}
P_{n_{total}}= \prod_{i=1}^{n_{total}-1}P_{i}\times P_{i+1}\times (1-\frac{T}{CT_{same}})
\end{eqnarray}

From Eq.(16), We have related probability of indirect communication with the probability of route establishment between the nodes. Hence, We have modeled communication time between nodes and related it with indirect communication between nodes. The speed of the vehicles/nodes have been incorporated in the communication as well as their transmission range.  When transmission range of the nodes are changed while keeping a particular constant value of their maximum and minimum speed, we see that communication time between nodes moving in the same direction but with different speed decreases with increase in transmission range.  The route establishment probability in turn decrease from eq. (15), which cause an decrease in the probability of indirect communication from eq. (16).

The route breakage probability of nodes moving in opposite direction given in [5] is; as $P_{opp}$
\begin{eqnarray}
P_{diff}=\frac{T}{CT_{diff}}
\end{eqnarray}

where, $T$ is the simulation time and $CT_{opp}$ is the average communication time of the nodes moving in opposite direction.
Let $ S=R/n $ be the separation distance between the nodes, where $R$ is the total length of the strip and $n$ is the total number of nodes on the strip as implied in [8]. The number of nodes moving in opposite direction within a transmission range of a certain node is given as $ N_{o} = \frac{2\times T_{r}}{S} $, where $T_{r}$ is the node's transmission range. The average lifetime of the links among the nodes that form route is given by:

\begin{eqnarray}
CT_{diff} = \frac{T_{diff_{s}} + T_{diff_{ds}}}{(V_{c}+1)^2}
\end{eqnarray}

where, $ T_{diff} $ is the communication time of the nodes moving in the opposite direction at the same speed and $ T_{opp_{ds}}$ is the communication time of the nodes moving in the opposite direction with different speed.

We can find the node's different speed same as in eq.(12). The total communication time of the nodes moving in the opposite direction with same speed can be found as:

\begin{eqnarray}
T_{comm_{opp_{s}}}=\frac{\sum_{i=j=0}^{V_{c}}\sum_{m=1}^{N_{o}}T}{N_{o}+1}
\end{eqnarray}

Similarly, when the nodes are moving in opposite direction with different speed, the total communication time between the nodes will be:

\tiny
\begin{eqnarray}
T_{comm_{opp_{ds}}}=\sum_{i=0}^{V_{c}}\sum_{j=0}^{V_{c}}\frac{\sum_{m=0}^{N_{o}}\frac{2\times T_{r}-md}{|{(V_{min}+i\Delta{V})-(V_{min}+j\Delta{V})}|}}{N_{o}+1}
\end{eqnarray}
\normalsize

The total probability of communication between nodes moving in opposite direction when link is established is then given by:
\begin{eqnarray}
P_{link}'=1-P_{opp}=1-\frac{T}{CT_{opp}}
\end{eqnarray}

The probability of indirect communication of nodes can also be related with the probability of route establishment between the nodes moving in opposite direction. The probability of route establishment decreases when the nodes are moving in opposite direction while increasing the $Range$. The probability of route breakage increases as nodes move in opposite direction with increase in transmission range.
The relationship can be shown as:

\begin{eqnarray}
P_{i}''=P_{i}\times P_{i}' \times P_{link}'=P_{i}\times P_{i}' \times (1-\frac{T}{CT_{opp}})
\end{eqnarray}

For total probability of indirect communication of $n$ segments as shown in eq.(9) we can write:

\begin{eqnarray}
P_{n_{total}}=\Pi_{i=1}^{n_{total}-1}P_{i}\times P_{i+1}\times P_{link}'
\end{eqnarray}

or
\begin{eqnarray}
P_{n_{total}}=\Pi_{i=1}^{n_{total}-1}P_{i}\times P_{i+1}\times (1-\frac{T}{CT_{opp}})
\end{eqnarray}

\section{Comparing Protocols with Throughput in NS-2}
In this paper, simulations are performed on two Mac layer protocols; 802.11 for MANETs and 802.11p for VANETs. We evaluate and compare the performance of three selected routing protocols by their default values; AODV, DSR and FSR and modified values; MOD AODV, MOD DSR and MOD FSR, with different scalabilities and varying mobilities in MANETs as well as in VANETs. we modify these chosen protocols and then analyze their results.

For MOD AODV, changes have been made in default AODV's ERS algorithm; $TTL\_START=2$, $TTL\_INCREMENT=4$ and $TTL\_THRESHOLD=9$. In MOD DSR, the modification made to original DSR consisted of reducing the size of Route Cache as taking $TAP\_CACHE\_SIZE=256$. Smaller Route Cache means that only relatively fresh routes are stored. In MOD FSR, intervals of inner and outer scopes of FSR are changed to $1s$ and $3s$ respectively.







Throughput is the measure of data received per unit time measured in bytes per second (bps). In MANETs, when throughput taken in scalability scenario.
Fig. 3.a,c shows that in smaller scalabilities, DSR has  highest throughput and FSR has the lowest throughput. This is because of route caching mechanism of DSR, whereas the scope routing of FSR is best suited for very large networks consisting of thousands of nodes. On the other hand, in higher scalabilities, AODV and MOD AODV both perform best and DSR has lowest throughput in MANETs. AODV provides more communication time as specified in equation because of the local link repair. Distance vector information broadcasting makes this protocol more apposite for denser networks.

In simulation results, Fig. 3.a,c depicts that MOD DSR gives better results in MANETs and in VANETs. We change cache size by decreasing its value to one fourth of its default value. By modifying $TAP\_ CACHE\_ SIZE$, fresher routes are available in the cache. As there is no explicit mechanism to delete stale routes in DSR, this modification helps to provide accurate routes as in equation (Probability of links due to fresher routes result more throughput).

In MANETs, MOD AODV not only improves its efficiency as compared to AODV but also outperforms among all other protocols. There is a less expanding ring values in initial default ERS values up to $TTL\_ THRESHOLD$. We expand these rings by increment the $TTL\_INCREMENT$ value from $2$ to $4$.It lessens routing delay and increase communication probability as from equation. Also this results, not only low routing load but also lowers the routing latency. Ultimately, the throughput value increased.

Overall, AODV and MOD AODV have the highest throughput values, comparing to all other protocols. LLR is a distinguish feature of AODV to make this protocol more scalable along with gratuitous RREPs.

\small
\vspace{0.05cm}

\begin{table}[!h]
\caption {Simulation Parameters}
\begin {center}
\begin{tabular}{|c|c|}
\hline
\textbf{Parameters} & \textbf{Values}\\
\hline

Simulator & NS-2(Version 2.34)\\

\hline

Channel type & Wireless  \\
\hline

Radio-propagation model  & Nakagami  \\
\hline

Network interface type & Phy/WirelessPhy, Phy/WirelessPhyExt  \\
 \hline

MAC Type  &Mac /802.11, Mac/802.11p  \\

\hline
Interface queue Type & Queue/DropTail/PriQueue \\
\hline

Bandwidth & 2Mb \\
\hline

Packet size & 512B  \\

\hline

Packet interval & 0.03s\\
\hline

 Number of mobile node & 25 nodes, 50 nodes, 75 nodes,100 nodes \\
\hline

Speed & 2 m/s,7 m/s,15 m/s,30 m/s\\
\hline

Traffic Type & UDP, CBR \\
\hline

Simulation Time & 900 s \\
\hline

Routing Protocols & AODV, DSR, FSR,  MOD AODV  \\
&  MOD DSR, MOD FSR \\
\hline
\end{tabular}
\end{center}
\end{table}

MOD DSR overall gives the highest throughput in high mobilities as compared to other protocols, while considering mobilities (Fig.3.b).
We observe that for low mobilites in MANETs MOD DSR outperform all other routing protocols; because DSR uses stale routes in frequent varying network topologies, on the other hand MOD DSR delete stale routes frequently as compared to DSR.
Therefore, MOD DSR due to reduced cache size makes DSR more efficient at low mobilites.
FSR due to absence of trigger updates performs worst among all shows the worst throughput as it is design for very dense network and therefore we also do not expect a high throughput from MOD FSR.

\begin{figure*}[t]
  \centering
 \subfigure[MANETs Throughput vs Scalability]{\includegraphics[height=5 cm,width=8 cm]{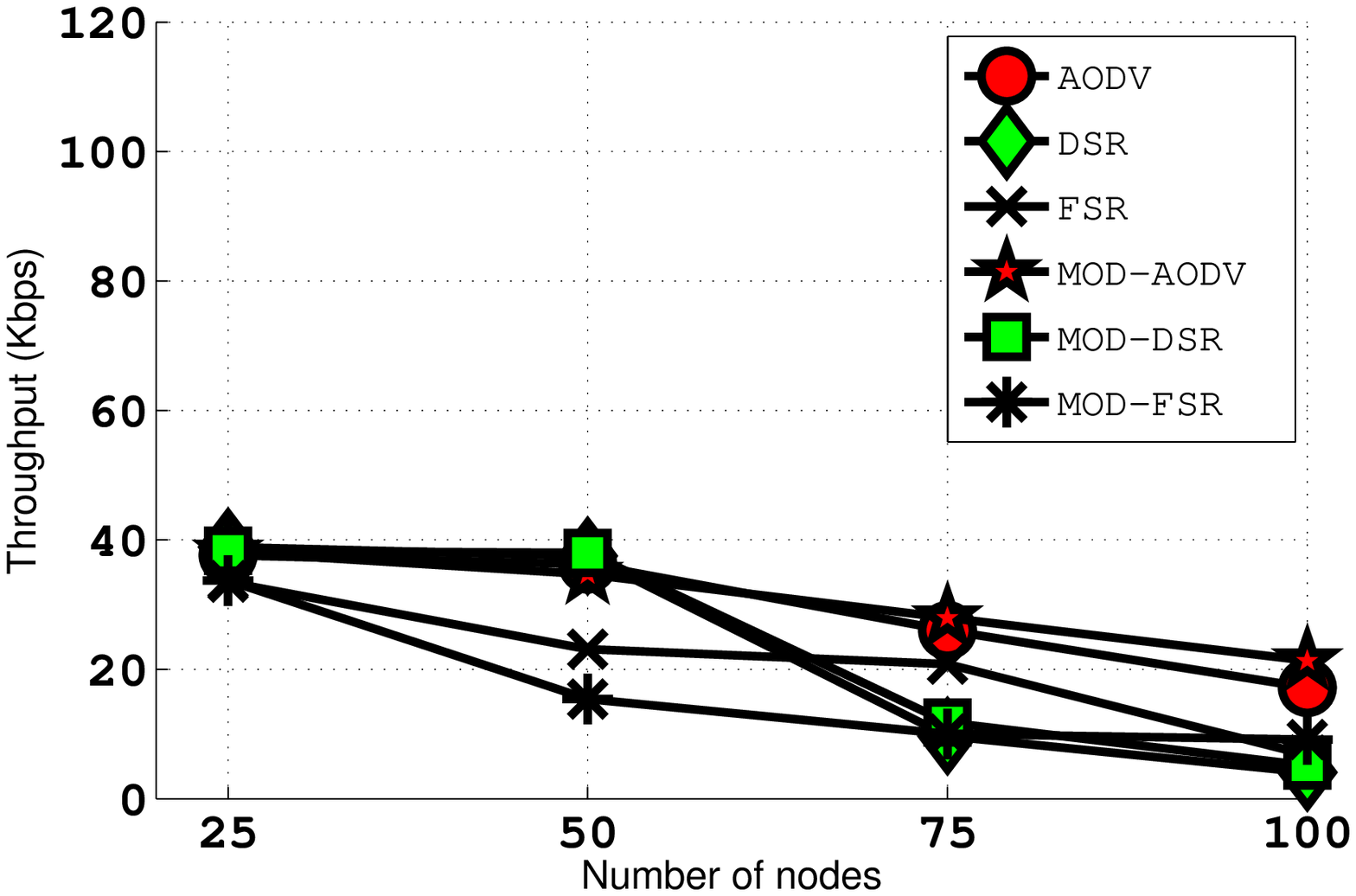}}
  \subfigure[MANETs Throughput vs Mobility]{\includegraphics[height=5  cm,width=8 cm]{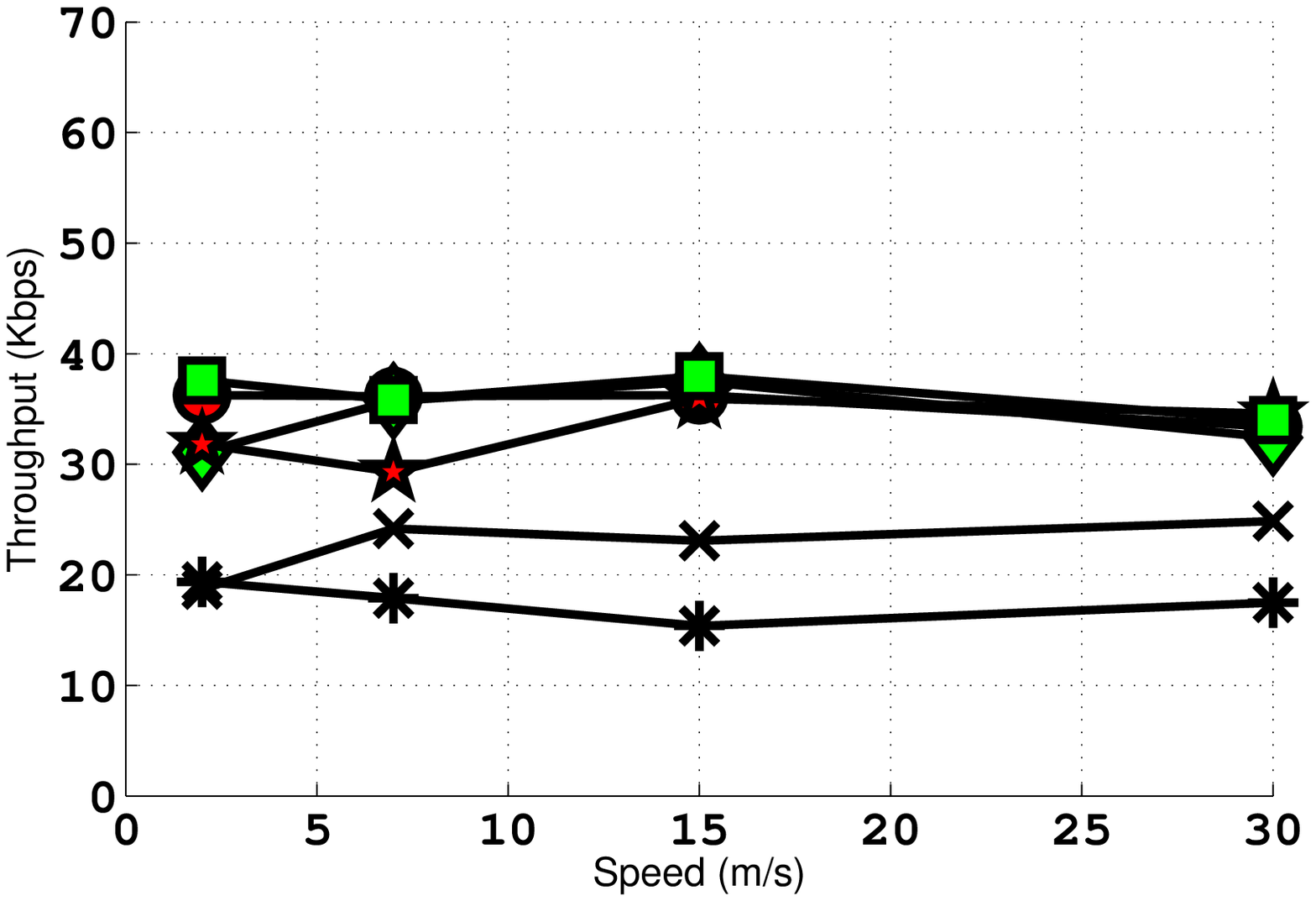}}
 \subfigure[VANETs Throughput vs Scalability ]{\includegraphics[height=5 cm,width=8 cm]{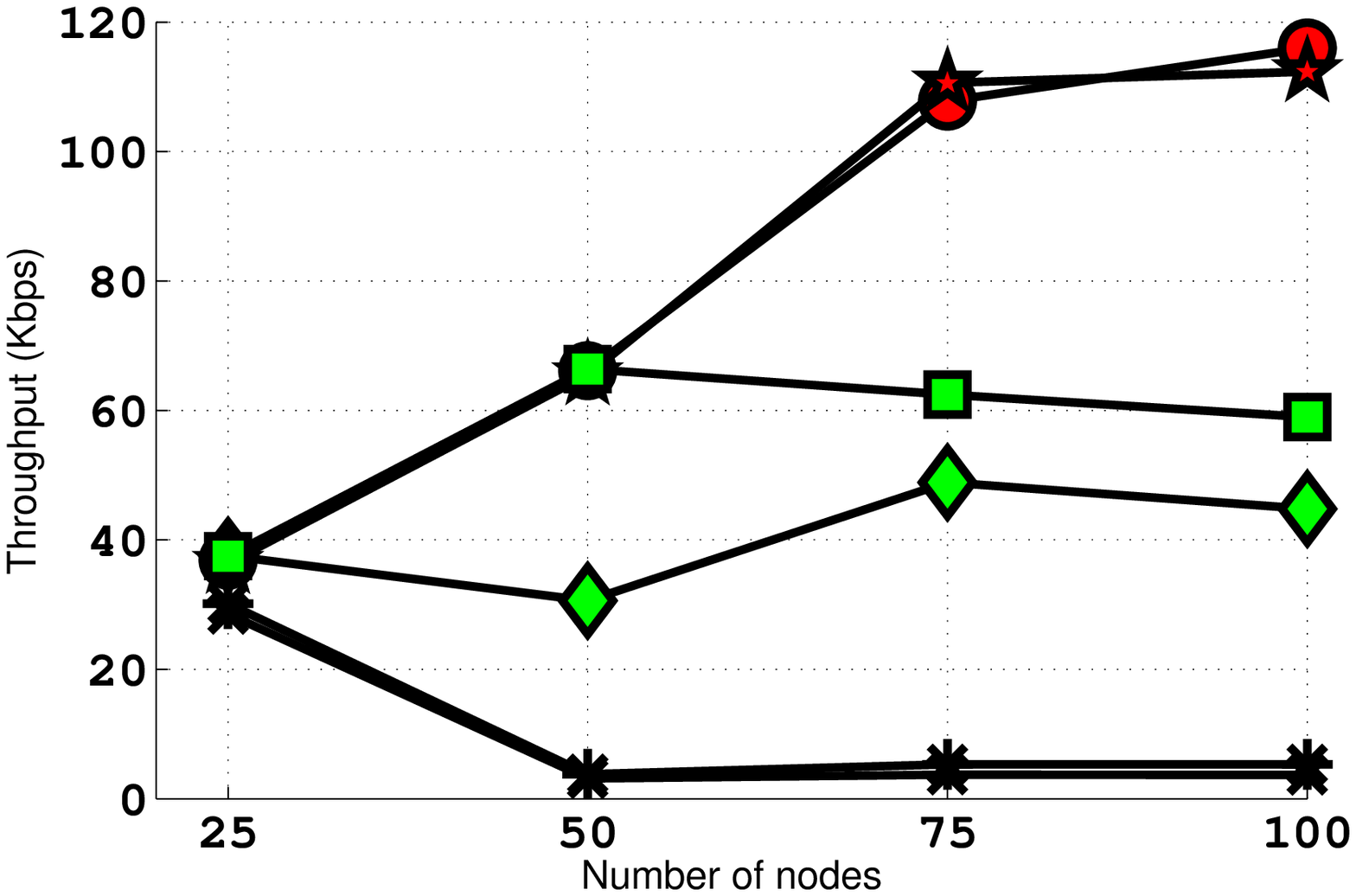}}
 \subfigure[VANETs Throughput vs Mobility ]{\includegraphics[height=5  cm,width=8 cm]{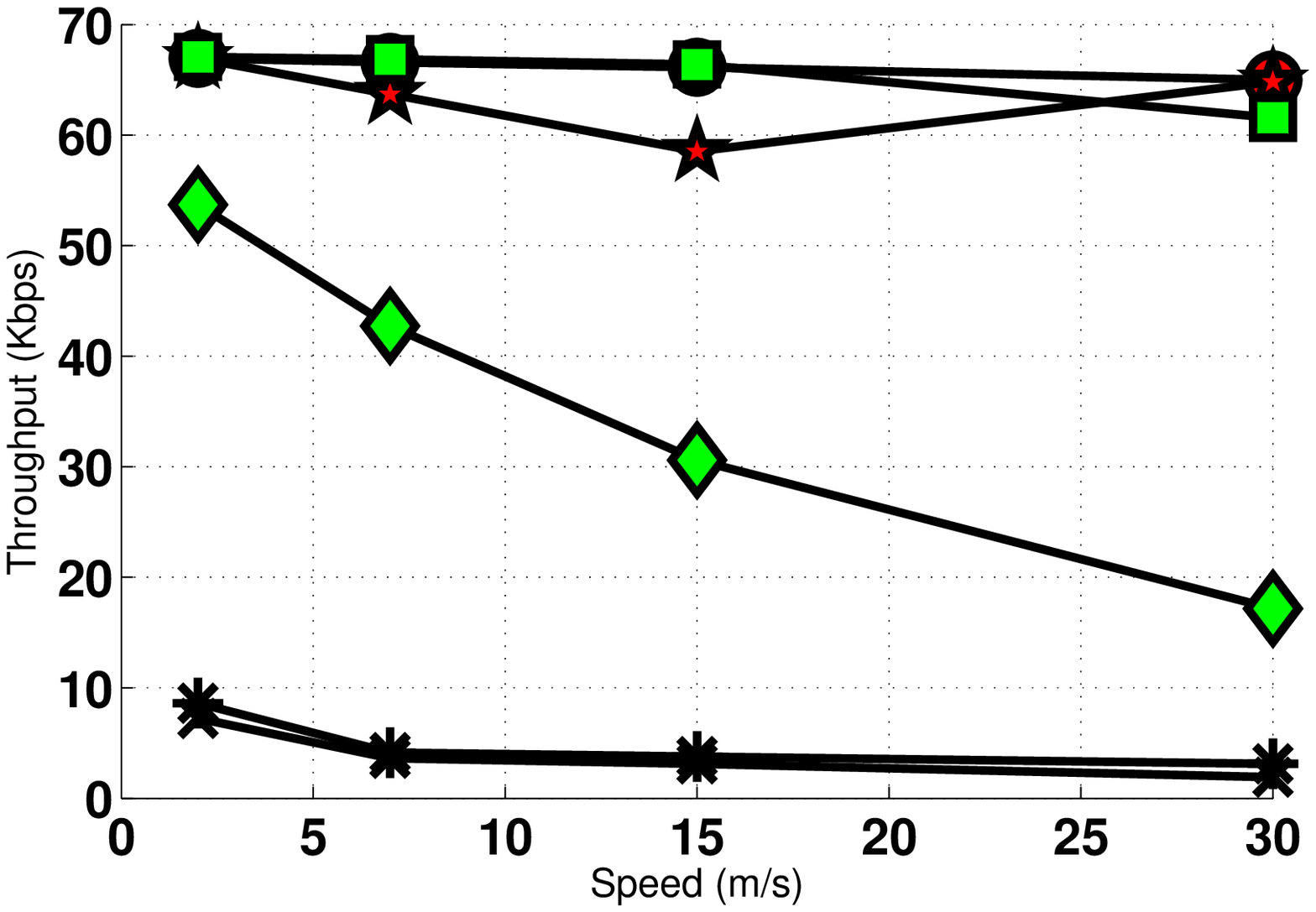}}
  \caption{Throughtput}
\end{figure*}
\vspace{0.3cm}
\hspace{0.3cm}\small

In high mobilities, link breakage is frequent, packet salvaging (PS) only efficient in moderate and in low dynamicity. To maintain the fresher routes route cache updating time interval must be shorten to avoid stale routes.
 After shortening the cache size, its efficiency becomes almost equal to original AODV in VANETs, as shown in Fig. 3.d. Grat. RREPs help DSR and AODV to have higher throughput in mobility scenario.

DSR has route cache mechanism which is not well suited for highly dynamic network topologies. FSR and MOD FSR are producing lowest results.
If we evaluate overall performance of all protocols in both MANETs and in VANETs then AODV which produces highest throughput, because it uses LLR, HELLO messages and gratuitous RREPs its advantage in highly mobile scenarios as depicted from  Fig. 3.b,d.

Here, as per our observation, throughput of MOD AODV in MANETs has shown almost similar behavior as for AODV; it decreased for scalability as expected but it worked well with highly mobile small network. For VANETs, throughput again shows similar behavior and decreased for both scenarios but change in throughput for mobility is minuscule while for scalability decreases is somewhat more prominent.

\section{Conclusion and Future Work}
In this paper, a framework is presented for node distribution with respect to density, network connectivity and communication time. Routing protocols DSR, AODV and FSR were compared in MANETs and VANETs. Besides evaluating the performance of AODV, DSR and FSR, we also made some modifications to these routing protocols and observed their performance. At the end we came up with the result that with minor changes, better results can be achieved in atleast one parameter. These changes provide better result. We conclude that AODV performs best among original protocols while MOD DSR produces highest throughput. In high speeds, DSR due to stale routes in route cache fails to converge. On the other hand in MOD DSR due to reduction in the size of route.

In future work, we are interested to implement Expected Transmission Count (ETX) link metric with mathematical modeling, as demonstrated in [11-13].

\small
\vspace{0.09cm}
\begin{table}{!h}
\caption {Performance Trade Offs by Routing Protocols}
\begin {center}
\begin{tabular}{|c|c|c|}

\hline
\textbf{Routing } & \textbf{Tadeoff} & \textbf{Resons} \\
\textbf{Protocols}&&\\
\hline

AODV 	&High Throughput at the & Local link repair, HELLO \\



 MOD 	  &Lower throughput at the &	  Smaller network diameter\\
AODV&cost of low E2ED and NRL&and less number of  \\
&&iterations of ring search \\
&&diameter\\
\hline

DSR 	&Average NRL and Average& Grat. RREPs, Source\\
&throughput at the cost of& routing and Packet \\
& high E2ED	&salvaging.\\
\hline

MOD& Higher throughput at the	& Smaller cache size and\\
 DSR&	 cost of low E2ED and NRL& availability of fresher \\
&&routes reduces the frequency\\
&& to compute a new route\\
\hline

FSR 	&Low E2ED at the cost &Graded-frequency technique\\
&of throughput and NRL.	& while using Fish-eye-state \\
&&routing algorithm\\
\hline
MOD &  Slightly better throughput &	Because of smaller interval \\
 FSR	&for low E2ED and more &of periodic updates \\
 &NRL &Routes are readily available\\
\hline

\end{tabular}
\end {center}
\end {table}

\end{document}